\documentclass[conference]{IEEEtran}
\IEEEoverridecommandlockouts
\usepackage[letterpaper, top=0.75in, bottom=1.04in, left=0.625in, right=0.625in, columnsep=0.25in]{geometry}
\usepackage{cite}
\usepackage{amsmath,amssymb,amsfonts}
\usepackage{algorithmic}
\usepackage{graphicx}
\usepackage{textcomp}
\usepackage{xcolor}
\usepackage[font=small,labelfont=bf]{caption}
\usepackage{subcaption}
\def\BibTeX{{\rm B\kern-.05em{\sc i\kern-.025em b}\kern-.08em
    T\kern-.1667em\lower.7ex\hbox{E}\kern-.125emX}}
\begin{document}

\title{Experimental Demonstration of Robust Distributed Wireless Clock Synchronization \\
\thanks{Efforts sponsored by the U.S. Government under the Training and Readiness Accelerator II (TReX II), OTA. The U.S. Government is authorized to reproduce and distribute reprints for Governmental purposes, notwithstanding any copyright notation thereon. The views and conclusions contained herein are those of the authors and should not be interpreted as necessarily representing the official policies or endorsements, either expressed or implied, of the U.S. Government.}
}

\author{\IEEEauthorblockN{
         Kumar Sai Bondada\IEEEauthorrefmark{1},
         Hiten Kothari\IEEEauthorrefmark{1},
         Yibin Liang\IEEEauthorrefmark{1},
         Daniel J. Jakubisin\IEEEauthorrefmark{1}\IEEEauthorrefmark{2},
         R. Michael Buehrer\IEEEauthorrefmark{1} \\
}
     \IEEEauthorblockA{
         \IEEEauthorrefmark{1}Wireless@VT, Bradley Department of ECE, Virginia Tech, Blacksburg, VA, USA \\
         \IEEEauthorrefmark{2}Virginia Tech National Security Institute, Blacksburg, VA, USA  
         }
           \vspace{-1cm}
         }

\maketitle
\begin{abstract}
Distributed wireless clock synchronization aligns the clocks of distributed transceivers wirelessly to support joint transmission and reception techniques. It allows for transceiver mobility and operates independently of GPS. A recently explored method involves synchronizing distributed transceivers using a two-tone waveform, in which the tones are separated in frequency by a clock (frequency) reference signal. Prior research has demonstrated frequency accuracy better than 1 Hz. However, this approach remains vulnerable to both intentional and unintentional interference. In this demonstration, we present a robust frequency-hopped two-tone waveform that enables transceivers to extract the reference signal for clock synchronization. 
\end{abstract}

\section{INTRODUCTION AND MOTIVATION}
Accurate synchronization in both frequency and time is crucial for coordinating distributed transceivers in next-generation wireless systems. In a typical transceiver, phase-locked loops (PLLs), referenced to a local crystal oscillator, generate the system clocks and carrier signals that drive RF chains and hardware components such as ADCs and DACs. In distributed architectures, however, each transceiver employs an independent oscillator that drifts over time, leading to clock and carrier mismatches. These offsets degrade system performance and hinder coherent functions such as joint transmission and reception.

While wired solutions such as the Octo-clock provide high frequency and timing accuracy, they inherently restrict system mobility. In contrast, wireless alternatives like GPS are limited by the high power required for clock extraction and by weak signal strength in many environments. Recent progress in research have introduced clock synchronization using a two-tone signal for frequency alignment and bi-directional waveform exchange for time alignment.

In this demo, we focus specifically on frequency synchronization. Extracting the clock signal from a two-tone waveform was first proposed in the AirShare paper \cite{7218555}, where the tones are spaced by the reference frequency required by transceivers in distributed systems. The two-tone signal is transmitted over the air and received by distributed transceivers, which extract the reference signal using specialized circuitry. A 10 MHz reference square wave derived from the two-tone signal is fed into the 10 MHz reference input of the USRP B210, which drives the PLL frequency synthesizer (ADF4001). This reference can also be converted into a pulse-per-second (PPS) signal using clock dividers, as demonstrated in the RFclock paper \cite{10.1145/3447993.3448623}. Subsequent studies have further examined and extended the use of two-tone waveforms. Readers are referred to the cited works for mathematical details and additional insights.

Although the two-tone waveform approach achieves sub-Hz frequency accuracy, it remains vulnerable to both intentional and unintentional interference. To address this limitation, we demonstrate a robust frequency-hopped two-tone waveform that allows transceivers to extract the reference signal without prior knowledge of the exact transmission frequency.

 \begin{figure}[htpb]
    \centering
    \begin{subfigure}{\columnwidth}
    \centering
    \includegraphics[width=\columnwidth]{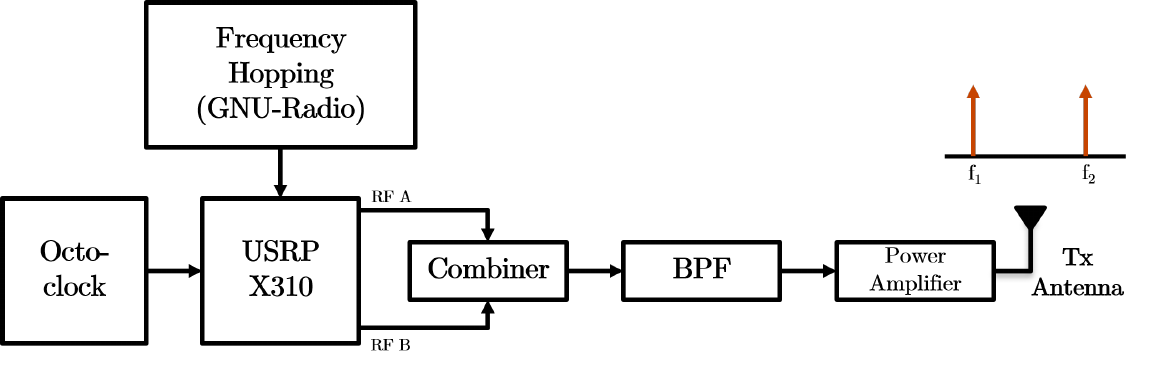}
    \caption{Two Tone generation at the leader}
    \label{fig:clock_gen1}
    \end{subfigure}
    \hfill
    \begin{subfigure}{\columnwidth}
    \centering
    \includegraphics[width=\columnwidth]{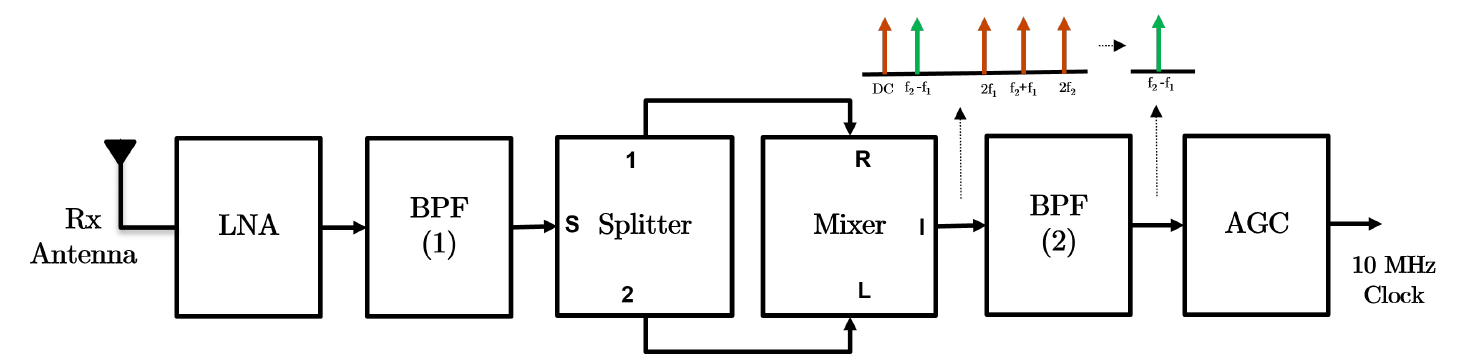}
    \caption{Extraction of clock reference signal at the follower}
    \label{fig:clock_gen2}
    \end{subfigure}
    \caption{Distributed clock synchronization circuit}
    \label{fig:freq_sync_circuit}
\end{figure}

\section{Demonstration Setup, Results, and Conclusion}
The proposed solution employs two-tone waveform hopping at the leader transceiver within the antenna-supported bandwidth, implemented in GNU Radio. The hopping is performed in the baseband to avoid RF retuning and the associated PLL settling delays. Specifically, the two tones hop around a frequency corresponding to the RF chains of the X310, with hopping executed in the baseband, which requires a large bandwidth. The availability of such bandwidth enables a larger number of frequency hops for hopping, providing improved flexibility in waveform design. An alternative approach is to hop the carriers directly, but this introduces delays due to PLL settling.

Follower transceivers operate without prior knowledge of the hopping pattern and require only a two-tone signal separated by the reference frequency. Two tones, \(f_1\) and \(f_2\), spaced 10~MHz apart, are generated using channels A and B of the X310, combined, amplified, and transmitted via an antenna, shown in Fig.~\ref{fig:clock_gen1}. The follower transceiver chain consists of a receive antenna, low-noise amplifier (LNA), bandpass filter (BPF) within the hopping bandwidth, splitter, combiner, a second 10~MHz BPF, and an automatic gain controller (AGC) to stabilize signal amplitude, as shown in Fig.~\ref{fig:clock_gen2}. The components are listed in Table~\ref{tab:components}. While the circuit can be further optimized, the primary goal is to extract a 10~MHz reference from the frequency-hopped two-tone waveform. The recovered clock is sinusoidal but can be converted to a square wave, which is optimal for USRPs, though they also accept sinusoidal signals.





\begin{table}[h]
\centering
\caption{RF Components and Their Frequency Ranges}
\begin{tabular}{|l|l|l|}
\hline
\textbf{Type}      & \textbf{Component}        & \textbf{Frequency Range (in MHz)}                  \\ \hline
Combiner  & ZFRSC–42                  & DC–4200                                \\ \hline
Splitter  & ZX10-2-12-s+                 & 2 - 1200                                \\ \hline

Mixer              & ZEM-4300                  & 300–4300                               \\ \hline
Filter             & SBP-10.7+          & 9.5 - 11.5       \\ \hline
Filter             & VBFZ-925-S+
          & 800-1050       \\ \hline
LNA                & ZX60-2534MA-S+                   & 500–2500                                \\ \hline
Antenna            & VERT900             & 824-960, 1710-1990                                           \\ \hline
AGC            & AD8367-EVALZ
             &   0-500                                         \\ \hline
\end{tabular}
\label{tab:components}
\end{table}

\begin{figure}[htpb]
    \centering
    \begin{subfigure}{\columnwidth}
    \centering
    \includegraphics[width=\columnwidth]{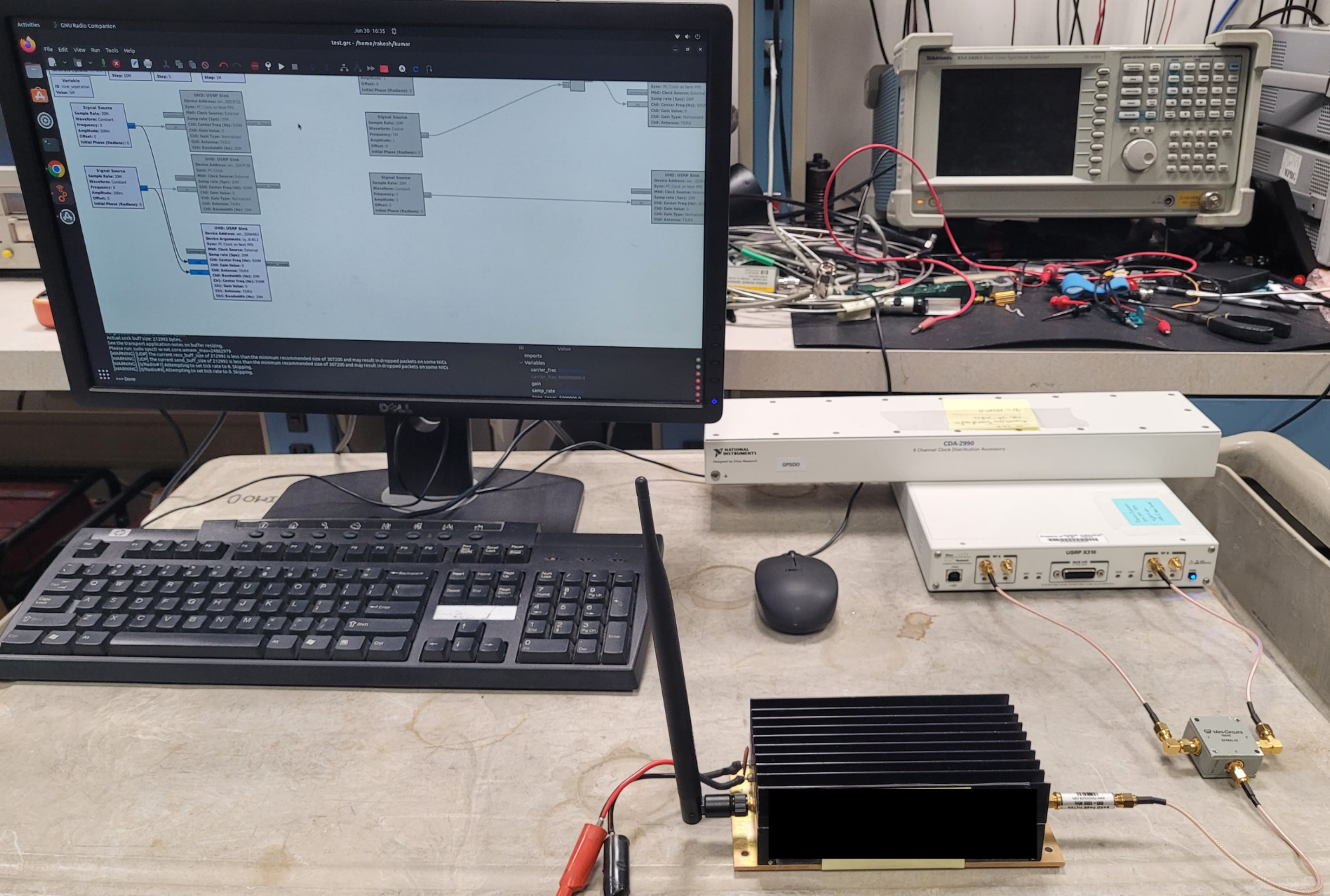}
\caption{Generation of a frequency-hopped two-tone waveform using GNU Radio and the X310 USRP. The tones $f_1$ and $f_2$ are generated at RF channels A and B, respectively, combined using a combiner, and passed through a power amplifier.}

    \label{fig:tx_circuit}
    \end{subfigure}
    \hfill
    \begin{subfigure}{\columnwidth}
    \centering
\includegraphics[width=\columnwidth]{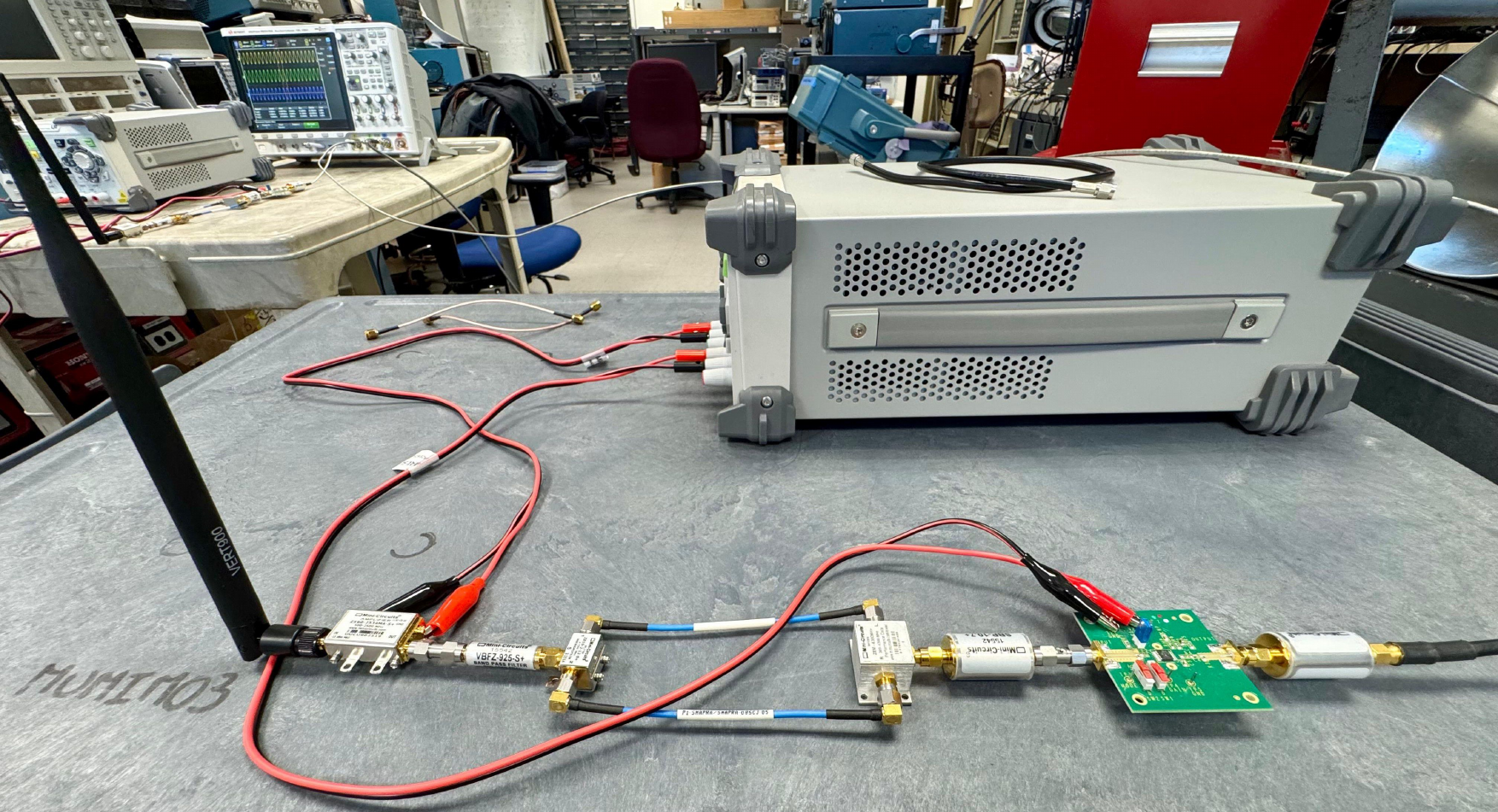}
\caption{Reference signal extraction circuit consisting of bandpass filters (BPF(1) at the reception frequency and BPF(2) at 10~MHz), LNA, splitter, combiner, and AGC.}

    \label{fig:rxCircuit_twotone}
    \end{subfigure}
    \caption{Distributed clock synchronization demonstration.}
    \label{fig:freq_sync_demo}
\end{figure}

\begin{figure}
    \centering
    \includegraphics[width=\columnwidth]{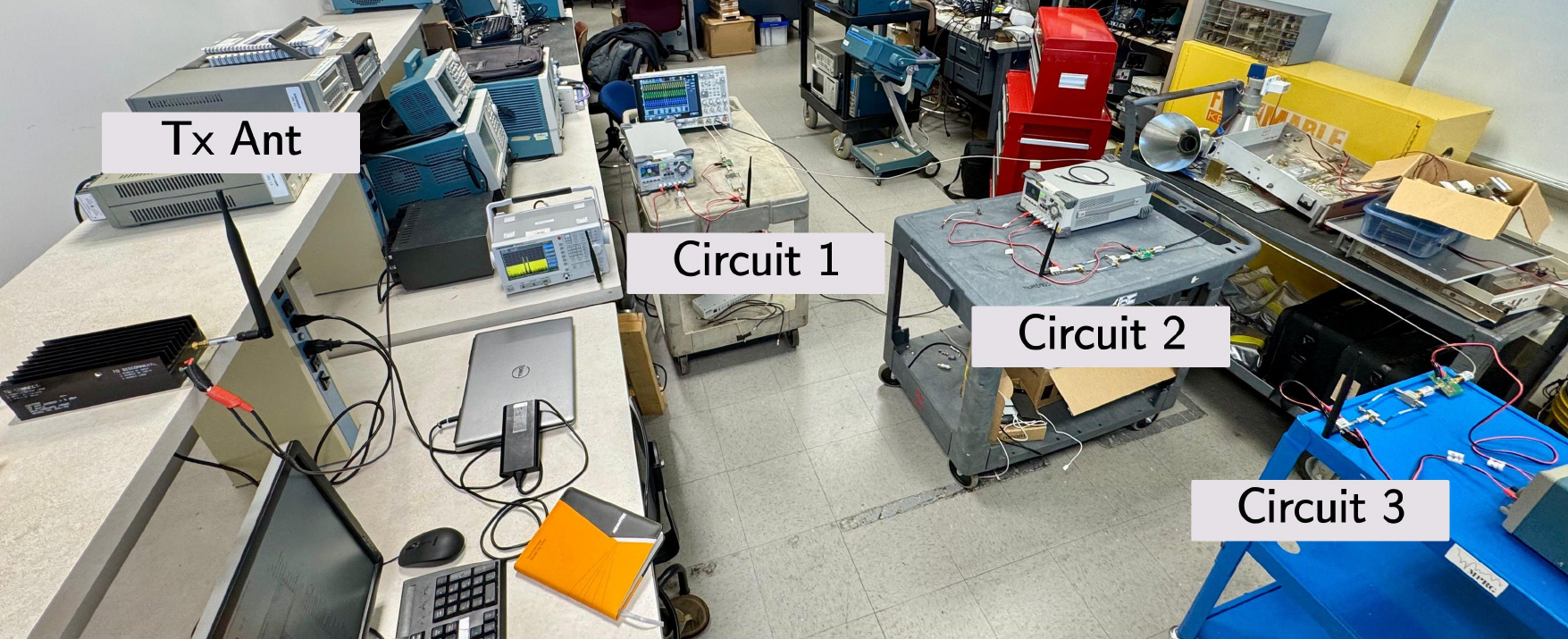}
\caption{Demonstration setup showing the leader transceiver and follower transceivers.}
    \label{fig:demo_setup-final}
\end{figure}

The demonstration setup is shown in Fig.~\ref{fig:freq_sync_demo}. The leader transceiver uses a power amplifier (PA) as the only active transmit-stage component, whereas the follower transceiver uses a low-noise amplifier (LNA) and automatic gain control (AGC) as its only active receive-stage components; all active components require an external voltage supply. The transceivers were placed 3~m apart from the leader transceiver, shown in Fig.~\ref{fig:demo_setup-final} and operated in the 900~MHz ISM band with periodic frequency hopping. The two-tone waveform hops randomly among center frequencies of 890, 895, 900, 905, and 910 MHz.

\begin{figure}
    \centering
    \includegraphics[width=\columnwidth]{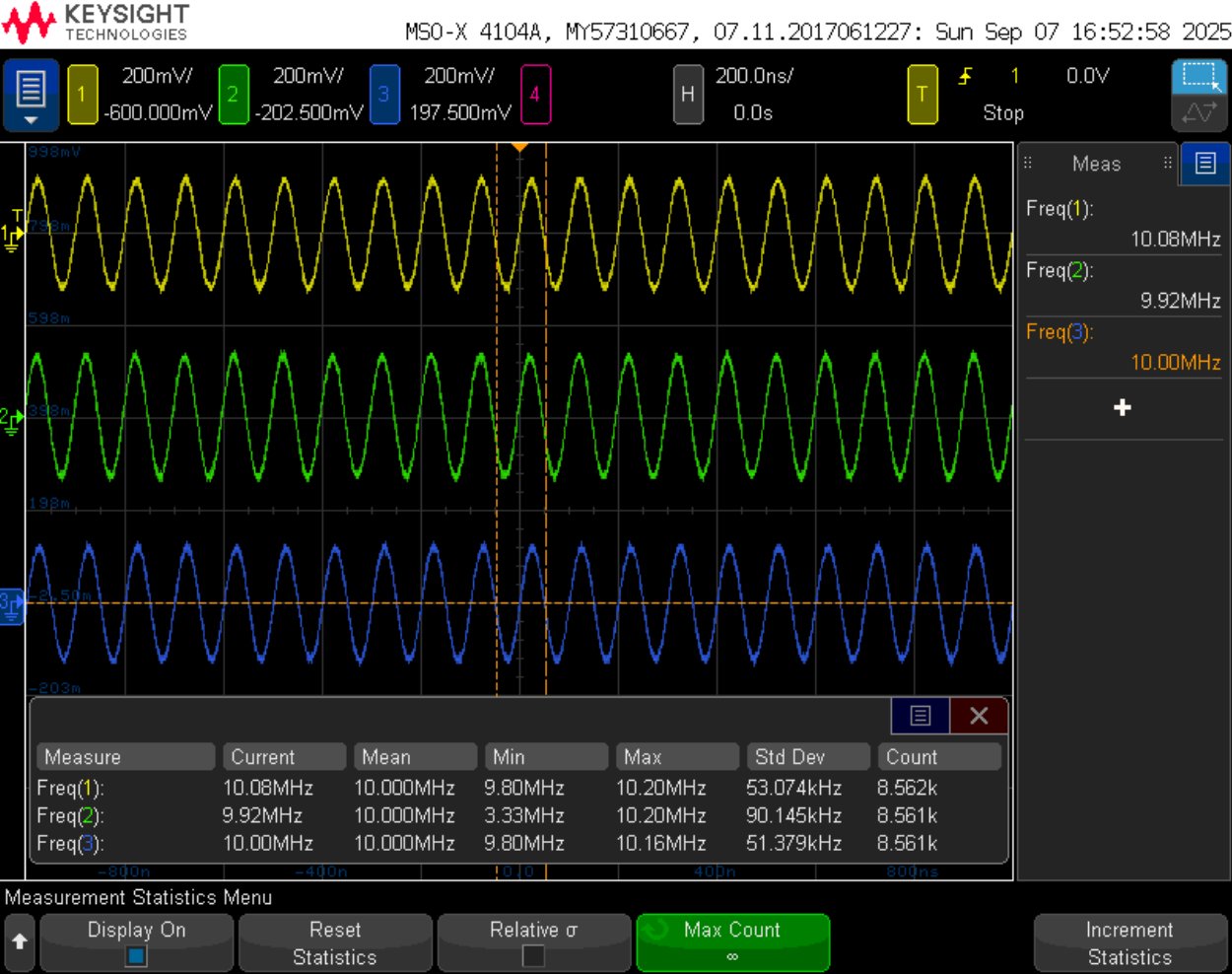}
\caption{Extracted 10 MHz reference clock signals displayed on the oscilloscope, showing the frequency statistics.}
    \label{fig:rxCircuitOutput-final}
\end{figure}

The circuit outputs are connected to an oscilloscope to observe the extracted clock signals over time, as shown in Fig.~\ref{fig:rxCircuitOutput-final}, together with a 10~MHz sinusoidal reference. All signals have a mean frequency of 10~MHz but are not time-aligned (i.e., not phase-synchronized). Time-synchronization techniques may be applied to achieve alignment~\cite{10.1145/3447993.3448623}. The nominal frequency-hopping interval is 1~second, while additional intervals of 0.5, 0.25, and 0.1~seconds are also used for testing.

\begin{figure}
    \centering
    \includegraphics[width=\columnwidth]{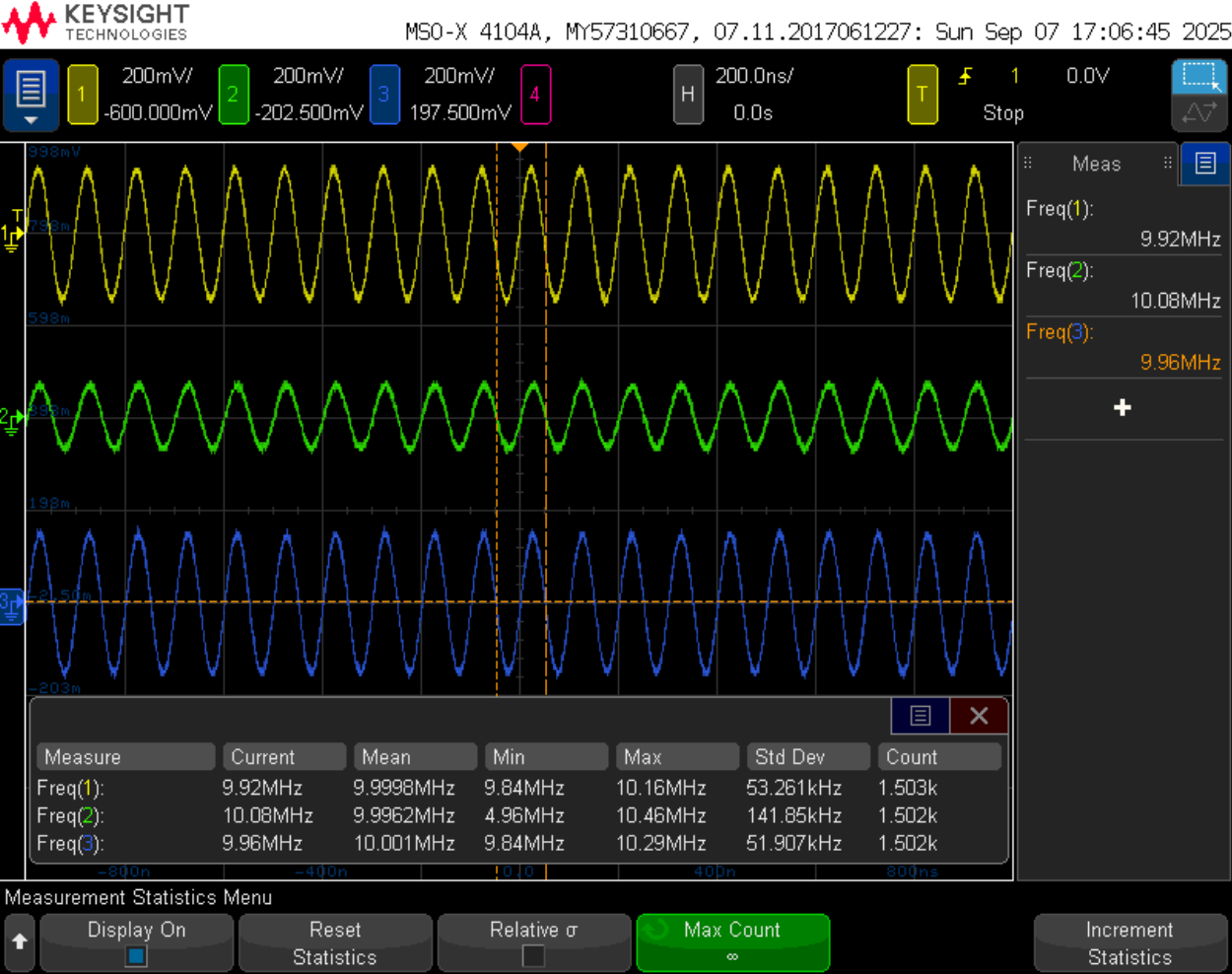}
\caption{Extracted 10 MHz reference clock signals displayed on the oscilloscope, showing the amplitude variation at the hopping instant.}
    \label{fig:rxCircuitOutput-hopping}
\end{figure}

The extracted 10~MHz reference clock signals were captured using an oscilloscope and analyzed offline in MATLAB. The phase of the extracted clock (sinusoid) can be expressed as
\begin{equation}
    \phi(t) = 2\pi f t + \phi_{0} + \omega(t),
\end{equation}
where $f$ denotes the frequency, $\phi_{0}$ the initial phase, and $\omega(t)$ the phase noise. The clock frequency $f$ was estimated using linear regression, and the extracted clocks remain near 10~MHz, with pairwise frequency differences among the follower transceivers on the order of $1$~Hz. Amplitude variations are observed at the hopping instants, as shown in Fig.~\ref{fig:rxCircuitOutput-hopping}. Consequently, as the hopping interval decreases, amplitude variations occur more frequently, which may disturb the clock signal. Future work will address the reduction of amplitude variations and utilize the extracted clocks to enable joint transmission and reception techniques, thereby demonstrating the effectiveness of the two-tone waveform.

\bibliographystyle{IEEEtran}
\bibliography{reference}
\end{document}